\documentclass[10pt,conference]{IEEEtran}
\IEEEoverridecommandlockouts

\usepackage{amsmath,amsfonts, amssymb}
\usepackage{algorithm}
\usepackage{adjustbox}
\usepackage{lscape}
\usepackage{siunitx}
\usepackage[noend]{algpseudocode}
\usepackage{textcomp}
\usepackage{fancyvrb}
\usepackage{graphicx}
\usepackage{soul}
\usepackage[many]{tcolorbox}
\usepackage{setspace}
\usepackage{booktabs}
\usepackage[utf8]{inputenc}
\usepackage[T1]{fontenc}
\usepackage{rotating}
\usepackage{graphicx}
\usepackage{paralist}
\usepackage{tabularx}
\usepackage{multicol}
\usepackage{multirow}
\usepackage{pbox}
\usepackage{enumitem}
\usepackage{colortbl}
\usepackage{pifont}
\usepackage{xspace}
\usepackage{url}
\usepackage{tikz}
\usepackage{tabularx}
\usepackage{fontawesome}
\usepackage{lscape}
\usepackage{listings}
\usepackage{color}
\usepackage{showexpl}
\usepackage{anyfontsize}
\usepackage{comment}
\usepackage{xtab}
\usepackage{soul}
\usepackage{multibib}
\usepackage{multirow}
\usepackage{xspace}
\usepackage{footnote}
\usepackage{tabularx}
\usepackage{tcolorbox}
\usepackage{booktabs}
\usepackage{balance}
\usepackage[normalem]{ulem}
\usepackage{nicematrix}
\usepackage{hyperref}

\usepackage{hhline}

\definecolor{main}{HTML}{5989cf}    
\definecolor{sub}{HTML}{cde4ff}     

\tcbset{
	sharp corners,
	colback = white,
	before skip = 0.2cm,    
	after skip = 0.5cm      
}                           

\newtcolorbox{boxM}{
	fontupper = \color{white},
	rounded corners,
	arc = 6pt,
	colback = main!80,
	colframe = main,
	boxrule = 0pt,
	bottomrule = 4.5pt,
	enhanced,
	fuzzy shadow = {0pt}{-3pt}{-0.5pt}{0.5pt}{black!35}
}

\newcolumntype{N}{>{\centering\arraybackslash}m{.85in}}
\def\BibTeX{{\rm B\kern-.05em{\sc i\kern-.025em b}\kern-.08em
    T\kern-.1667em\lower.7ex\hbox{E}\kern-.125emX}}

\newboolean{showcomments}

\setboolean{showcomments}{true}

\ifthenelse{\boolean{showcomments}}
{\newcommand{\nb}[2]{
		\fbox{\bfseries\sffamily\scriptsize#1}
		{\sf\small$\blacktriangleright$\textit{#2}$\blacktriangleleft$}
	}
	
}
{\newcommand{\nb}[2]{}
	
}

\newcommand{\ie}{\emph{i.e.,}\xspace}
\newcommand{\eg}{\emph{e.g.,}\xspace}
\newcommand{\etc}{etc.\xspace}
\newcommand{\etal}{\emph{et~al.}\xspace}
\newcommand{\secref}[1]{Section~\ref{#1}\xspace}
\newcommand{\figref}[1]{Fig.~\ref{#1}\xspace}

\newcommand{\tabref}[1]{Table~\ref{#1}\xspace}

\newcommand{\RQ}[1]{RQ$_{#1}$\xspace}
\newcommand{\databw}{$D_{b\&w}$\xspace}
\newcommand{\datad}{$D_{d}$\xspace}
\newcommand{\datas}{$D_{s}$\xspace}

\definecolor{lightergray}{rgb}{0.9,0.9,0.9}
\newtcolorbox{resultbox}{colback=lightergray, arc=0.5mm, top=2mm, bottom=2mm, left=2mm, right=2mm}

\definecolor{arsenic}{rgb}{0.23, 0.27, 0.29}
\definecolor{darkgray}{rgb}{0.33, 0.33, 0.33}

\definecolor{Gray}{gray}{0.9}
\definecolor{codegreen}{rgb}{0.44,0.57,0.55}
\definecolor{codegray}{rgb}{0.73,0.38,0.06}
\definecolor{codepurple}{rgb}{0.70,0.27,0}
\definecolor{codemagenta}{rgb}{0.74,0.09,0.42}
\definecolor{codeoutput}{rgb}{0.5,0,0}
\definecolor{backcolour}{rgb}{0.98,0.98,0.98}

\definecolor{codekone}{rgb}{0.20,0.50,0.20}
\definecolor{codektwo}{rgb}{0.66,0.02,0.24}
\definecolor{codekthree}{rgb}{0.58,0.20,0.80}
\definecolor{codekfour}{rgb}{0.58,0.20,0.80}

\lstdefinestyle{mystyle}{
    backgroundcolor=\color{backcolour},
    commentstyle=\color{codegreen},
    numberstyle=\tiny\color{codegray},
    stringstyle=\color{codemagenta},
    keywordstyle=\color{codekone},
    language=Java,
    breakatwhitespace=false,         
    breaklines=true,                 
    keepspaces=true,                 
    numbers=none,                    
    numbersep=5pt,                  
    showspaces=false,                
    showstringspaces=false,
    showtabs=false,                  
    tabsize=2,
    frame=tb,
    framerule=0pt,
    basicstyle=\fontsize{7}{7}\fontfamily{\ttdefault}\selectfont,
    moredelim=[is][\textcolor{codektwo}]{???}{???},
    moredelim=[is][\textcolor{codekthree}]{\%\%}{\%\%}
}

\begin{document}
\addtolength{\extrarowheight}{\belowrulesep}
\aboverulesep=0pt
\belowrulesep=0pt
\title{Personalized Code Readability Assessment: \\Are We There Yet?}

\author{
\IEEEauthorblockN{Antonio Vitale\IEEEauthorrefmark{1}\IEEEauthorrefmark{2}, Emanuela Guglielmi\IEEEauthorrefmark{1}, Rocco Oliveto\IEEEauthorrefmark{1}, and Simone Scalabrino\IEEEauthorrefmark{1}}
\IEEEauthorblockA{\IEEEauthorrefmark{1}DEVISER @ University of Molise, Italy, \{emanuela.guglielmi, rocco.oliveto, simone.scalabrino\}@unimol.it}
\IEEEauthorblockA{\IEEEauthorrefmark{2}Politecnico di Torino, Italy, antonio.vitale@polito.it}
}

\maketitle

\begin{abstract}
Unreadable code could be a breeding ground for errors. Thus, previous work defined approaches based on machine learning to automatically assess code readability that can warn developers when some code artifacts (\eg classes) become unreadable. Given datasets of code snippets manually evaluated by several developers in terms of their perceived readability, such approaches (i) establish a snippet-level ground truth, and (ii) train a binary (readable/unreadable) or a ternary (readable/neutral/unreadable) code readability classifier. Given this procedure, all existing approaches neglect the subjectiveness of code readability, \ie the possible different developer-specific nuances in the code readability perception.
In this paper, we aim to understand to what extent it is possible to assess code readability as subjectively perceived by developers through a \textit{personalized} code readability assessment approach. This problem is significantly more challenging than the snippet-level classification problem: We assume that, in a realistic scenario, a given developer is keen to provide only a few code readability evaluations, thus less data is available. For this reason, we adopt an LLM with few-shot learning to achieve our goal. Our results, however, show that such an approach achieves worse results than a state-of-the-art feature-based model that is trained to work at the snippet-level. We tried to understand why this happens by looking more closely at the quality of the available code readability datasets and assessed the consistency of the inter-developer evaluations. We observed that up to a third of the evaluations are self-contradictory. Our negative results call for new and more reliable code readability datasets.
\end{abstract}


\section{Introduction} \label{sec:intro}

Developers read code all the time. They might do that to fix a bug, to improve an existing code base, or even while they write code from scratch if they need to evaluate the code provided by a coding assistant such as Copilot.
Indeed, as research shows, code reading and understanding is the most common activity made by software developers \cite{minelli2015know}. 
Therefore, code readability, \ie the ease with which a program can convey information to a reader, is a particularly desirable property of the source code.
Unreadable code can both make developers struggle more while they try to acquire information from the source code (which implies higher software maintenance costs in the long run) and could better hide bugs.
Automatically finding unreadable code in large code bases is important for several reasons. Practitioners could use this information to decide where they should focus their effort in code cleaning and improvement activities. Researchers, on the other hand, can study how readability (or lack thereof) impacts other software properties \cite{scalabrino2017automatically, scalabrino2019automatically} or activities \cite{pantiuchina2020developers}.

Given the potential importance of automatically assessing code readability, previous work focused on defining approaches to achieve this goal. Such approaches are based on Machine Learning (ML) or Deep Learning (DL) models and, thus, they need to be trained on a labeled dataset. 
Earlier ML-based approaches \cite{buse2009learning, posnett2011simpler, dorn2012general, scalabrino2018comprehensive} relied on feature engineering. They defined structural, visual, and textual features that can be measured on the a given source code snippet. Those features are then used to characterize the (labeled) source code. Finally, such datasets are adopted to train a binary classifier that can categorize code snippets as ``readable'' or ``unreadable.'' 
Recent literature explored the possibility of adopting DL-based approaches \cite{mi2018improving, mi2022towards} to improve the accuracy of existing models. In this case, it is not necessary to manually define the features: The source code is treated as either an image \cite{mi2018improving, mi2022towards} or text \cite{mi2022towards}.

To the best of our knowledge, the literature provides three labeled code readability datasets, including single developer assessments on a non-binary scale, for training and evaluating readability assessment approaches \cite{buse2009learning, dorn2012general, scalabrino2016improving}. All of them are defined using the same core procedure. Several developers were asked to manually assess the readability of a given snippets using a 1 to 5 Likert scale, where 1 indicates that the snippet is completely unreadable, while 5 indicates that the snippet is completely readable.
Then, given a snippet, all the evaluation it received are combined (\eg by computing their average) to define the snippet-level code readability score ground truth. Such values are used to determine the labels, which might be two (\textit{readable} or \textit{unreadable}) or three (\textit{readable}, \textit{unreadable}, or \textit{neutral}). The mapping is based on the scores distribution, \eg the average score could be used as a threshold \cite{dorn2012general, scalabrino2016improving, scalabrino2018comprehensive}.

All previous work focused on evaluating code readability as a property of the \textit{source code} alone (\textit{generalist} code readability prediction). 
Let us go back to the definition of code readability we gave before, \ie readability is the ease with which a program can convey information to a reader. Code readability depends not only from the \textit{program}, but also from the \textit{reader}. Different developers might find code written in different ways more or less readable. 
\begin{figure}
    \centering
    \includegraphics[width=\linewidth]{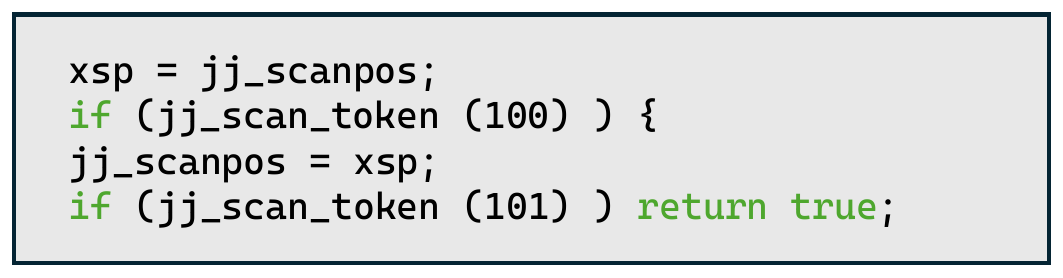}
    \caption{Example of controversial snippet.}
    \label{fig:intro:example}
\end{figure}
Let us consider the example in \figref{fig:intro:example} taken from the dataset by Buse and Weimer \cite{buse2008metric}. Some developers marked this snippet as \textit{highly readable} (5/5), while others as \textit{highly unreadable} (1/5). Probably, some evaluators focused more on its conciseness (thus marking it as readable), while others on the lack of indentation or on the lowly descriptive identifiers (which make it unreadable to some).

Despite the variability of the judgments of single developers, no previous work has tried to define a \textit{personalized} code readability assessment model, \ie a model that is based not only on the characteristics of the source code, but also on the preferences of the reader. Such a model might be useful to better allocate the resources in maintenance activities, \eg to reduce the chances that developers find themselves maintaining (and thus reading) code they will struggle with.

In this paper, we present an empirical study in which we aim to understand whether it is possible to define a \textit{personalized} code readability model.
The main challenge we faced while trying to define such a model lied in the lack of training data: A new reader would need to manually provide their own code readability evaluations to the train the model. We assume that, in practice, developers would accept to provide only a few of such evaluations. The only viable solution we saw fit to solve this problem was a Large Language Model (LLM). LLMs are pre-trained to achieve a complete knowledge of several programming languages. Thus, as previous work shows, a few examples are sufficient to train the model to complete a given task (few-shot learning).

First, we studied how \textit{generalist} code readability assessment models perform in the task of assessing readability at developer level. We compared a state-of-the-art model (the one by Scalabrino \etal \cite{scalabrino2018comprehensive}) with an additional baseline that relies on an LLM (GPT-4o) and with an ideal generalist approach which predicts, for a given snippet, the readability as perceived by the majority of developers who evaluated it. We found that both the realistic approaches achieve result which are close to the ideal model. Still, in any case, we observe that generalist models can predict developer-specific readability judgments with a relatively small accuracy (64\% and 47\%, depending on the dataset).
Thus, we compare the best (realistic) generalist model with an LLM-based personalized model. Surprisingly, the personalized model achieves worse result than the generalist model.

To analyze this phenomenon more in depth, we performed a manual evaluation of the datasets we adopted in our experiments. Specifically, we tried to look for evidence of self-contradictory evaluations, \ie readability evaluations given to two different snippets by the same developers that contradict each other (\ie we could find no rational explanation in the code that motivated the evaluations). We found that about 32\% and 23\% of the evaluations in the datasets by Buse and Weimer \cite{buse2008metric} and Scalabrino \etal \cite{scalabrino2016improving} might contain contradictions, respectively.
Our results call for the for the definition of new labeled dataset with a more principled and objective way of annotating the snippets in terms of readability. Also, our study shows that there is a substantial margin of improvement obtainable from a personalized model as for the evaluations provided by specific developers, we are still far from having such models.

\section{Background: Code Readability Assessment}
\label{sec:background}
In this section, we first introduce the code readability datasets available in the literature. Then, we present the state-of-the-art \textit{generalist} code readability assessment models. Finally, we conceptualize \textit{personalized} code readability assessment.

\subsection{Code Readability Datasets}
A code readability dataset consists of a set of \textit{code snippets}. Each snippet is associated with a set of readability assessments provided by \textit{developers}. There are three main datasets in the literature.

The one by Buse and Weimer (\databw) \cite{buse2008metric, buse2009learning} contains 100 Java snippets, each of them evaluated by 120 developers. The authors selected snippets made of three consecutive simple statements (\eg assignments, function calls) with other context such as comments, excluding trivial snippets (\eg import statements, only comments, \etc). Therefore, the snippets are relatively short and some of them are not compilable (\eg they lack closing brackets, such as in the example in \figref{fig:intro:example}). The annotators that volunteered for the study were primarily computer science students.
Thus, \databw is a dense 100x120 matrix.

The dataset by Dorn (\datad) \cite{dorn2012general} contains 360 snippets in Java, Python, and CUDA, each of which has been evaluated by a diverse subset of the developers. In other words, while 5,468 developers evaluated the snippet, not all developers evaluated all the snippets. Thus, \datad is a sparse 360x5,468 matrix.

Finally, the dataset by Scalabrino \etal (\datas)\cite{scalabrino2016improving} contains 200 Java snippet, all of which have been evaluated by 9 developers. Specifically, the authors first collected the methods from 4 open-source Java projects, then selected the 200 most representative snippets to enrich the diversity of the collected data. Differently from the other readability datasets, the collected snippets have a method-granularity level. Such snippets were rated by 9 computer science students.
\datas is a dense 200x9 matrix.

In all such datasets, readability has been assessed by asking developers to provide a subjective evaluation on a Likert scale from 1 (highly unreadable) to 5 (highly readable). Thus, all the elements of \databw, \datad, and \datas are in the set $\{1, 2, 3, 4, 5\}$.

\subsection{Generalist Code Readability Assessment Models}
Buse and Weimer \cite{buse2008metric,buse2009learning} were the first to introduce a model that can automatically assess code readability. They introduced a set of \textit{structural} features, such as line length and number of identifiers and used them to train a binary classifier. They trained and tested their model on \databw. To do that, they assigned a ground-truth readability score to each snippet by averaging the evaluations provided by the developers for it. They set a threshold for labeling the snippets as \textit{readable} and \textit{unreadable} based on the natural bimodal distribution of the resulting readability scores of the snippets.

Posnett \etal \cite{posnett2011simpler} proposed a simpler model for code readability that relied on only three metrics: Halstead volume, token entropy, and lines of code (LOC). This approach outperformed the earlier model, showing that fewer, well-selected features could effectively model code readability. They evaluated their approach on \databw using the same methodology adopted by Buse and Weimer \cite{buse2008metric} for assigning the binary labels to the snippets.

Dorn \cite{dorn2012general} addressed the limitations of earlier Java-specific models by introducing a more general approach applicable to multiple programming languages. Dorn's model incorporates a broader range of features, including new structural features, visual features, and a few textual features. Dorn identified seven features that most strongly correlated with human readability evaluation, achieving a Spearman correlation of 0.72. He tested his model on \datad. Also in this case, he first computed the snippet-level readability scores by average the developers' evaluation. Given that the distribution of ground-truth readability scores in his dataset did not follow a bimodal distribution (differently from \databw), he used the average score as a threshold.
His model outperformed the retrained model by Buse and Weimer \cite{buse2009learning}.

Scalabrino \etal \cite{scalabrino2016improving, scalabrino2018comprehensive} introduced a comprehensive model that integrates structural, visual, and several new textual features. They compared such a model with the previously-mentioned ones on all the datasets, \ie \databw, \datad, and \datas. As for the latter, they used the same methodology adopted by Dorn \cite{dorn2012general} to set a threshold.

Mi \etal tried for the first time to adopt Deep-Learning-based approachs by means of a Convolutional Neural Network (CNN) \cite{mi2018improving}, graph neural networks \cite{mi2023graph}, and a combination of representations and techniques \cite{mi2022towards} to capture semantic, visual, and semantic aspects of the source code. Differently from previous work, they aimed at classifying snippets as \textit{readable}, \textit{unreadable}, or \textit{neutral} (\ie ambiguous). To do that, they computed for each snippet in \databw, \datad, and \datas the average readability score. Then, based on the distribution of the scores, they assigned the label \textit{unreadable} to the first quartile, the label \textit{readable} to the fourth quartile, and the label \textit{neutral} to the remaining 50\% of the snippets.

\subsection{Personalized Code Readability Assessment}
The basic step behind all the generalist code readability assessment approaches consists of aggregating the evaluations provided by developers to have a ground-truth readability score for a given snippet. Conversely, a \textit{personalized} code readability assessment approach does not require this step. In other words, the task consists of predicting the specific readability evaluation that a given developer would provide for a given snippet. 

It is worth noting that this change of perspective comes at a price. A generalist model can rely on the (snippet-based) readability data available on several snippets (\eg all the ones available in \databw, \datad, and \datad) to predict the readability of a new unlabeled snippet, regardless of the developer who needs this information. On the other hand, a personalized model requires some training data that specifically comes from the developer $d$ who is willing to get the recommendation. In other words, the datasets available can not be adopted (alone) to define a personalized model in practice for developers different from the ones who contributed to build them.

If we want to simulate such a scenario with the available datasets, we need to re-define the learning base that can be adopted to train a personalized model. Given any dataset $D_*$ (assuming a dense snippet-developers matrix), a personalized model can only learn from $D^{-d}_{*}$, \ie a version of $D_*$ from which the column of the evaluations provided by the developer $d$ is removed. Besides that, the model can rely on a very small number of evaluations from the evaluation column of developer $d$ (\ie the ones that $d$ would be asked to provide to train the model).

In this context, the classification label is not based on an aggregation, but rather on the original score provided by the developer (\ie the evaluation on a 5-points Likert scale, in the currently available datasets). The choice of the threshold for assigning a label should be strictly related to the methodology adopted to acquire the assessments. Let us assume that a Likert scale is used since all available datasets adopt such a methodology to operationalize readability. If the objective is, for example, to define a binary classification approach focused on identifying snippets that a developer might struggle with, an option could be to consider $\{1, 2\}$ as clear indications of the fact that the snippet is ``unreadable'' and $\{3, 4, 5\}$ as indications that the snippet does not have substantial readability problems (\ie ``readable'').

To tackle the challenge related to the small number of developer-specific data that can be used to train the model, we propose the use of an LLM. LLMs are pre-trained on very large codebases so that they learn the basics of the programming language(s). Such models can be trained with just a few examples (few-shot/in-context learning). 
Another problem that arises in this context is the choice of the examples that should be used for few-shot learning. Ideally, such examples should reflect the peculiarities of the developer $d$. Remember that we assume we do not have any code readability evaluation by $d$ and we want to simulate the scenario in which we choose which snippet we should ask them to evaluate before they can use the model.

We report in the next section the snippet selection algorithms we tested in our empirical investigation for selecting $k$ best snippets from the available learning base (\ie the evaluations provided by \textit{other} developers, such as the ones available in \databw, \datad, and \datas).

\section{Empirical Study Design}
\label{sec:design}
The goal of our study is to understand whether personalized code readability assessment is feasible.

Our study is steered by the following Research Questions (RQs):
\begin{itemize}
\item{\RQ{1}}: \textit{How effective are generalist models in predicting developers' subjective assessments of code readability?} With this first research question, we aim to understand if personalized models are needed in the first place by assessing whether generalist models are good enough in this different evaluation scenario.
\item{\RQ{2}}: \textit{How effective is an LLM-based personalized code readability assessment model in predicting developers subjective assessments of code readability?} With the second research question, we want to compare the best performing generalist model from the previous RQ with three versions of an LLM-based personalized model we devised.
\item{\RQ{3}}: \textit{To what extent are developers' code readability assessments consistent?} As we will show later, we obtained a negative result for \RQ{2}. With this last research question, we want to investigate more in-depth the reasons why this happened by studying the consistency of the readability evaluations in the datasets we adopted.
\end{itemize}

\subsection{Study Context}
The context of the study consists of Java snippets and the related developers readability assessments. We decided to adopt the dataset by Buse and Weimer \cite{buse2009learning} and the one by Scalabrino \etal \cite{scalabrino2016improving}.
As previously reported in \secref{sec:background}, the dataset by Buse and Weimer \cite{buse2009learning} (\databw), contains the evaluations of 120 human annotators which assessed the readability score of 100 Java code snippets, for a total of 12,000 annotations, while the dataset by Scalabrino \etal \cite{scalabrino2016improving} (\datas), contains the evaluations of 200 Java code snippets by 9 evaluators, totaling 1,800 annotations. 

We opted for \databw and \datas because (i) they contain complete evaluations from all annotators for each snippet (\ie they are dense matrices, differently from \datad), and (ii) they feature diverse snippet sizes (both small --- \databw --- and medium --- \datas). These factors allow us for a more detailed analysis of the singular developer assessments and for more generalizable findings across the different levels of code granularity.

As previously explained, since we aim to tackle the problem of predicting developer-specific code readability evaluations, we do not need to aggregate the evaluations for each snippet, differently from previous work. Instead, we need to define the labels we will need to predict and a function to map the readability scores assigned by the developers to such labels. We decided to adopt three labels as in recent work \cite{mi2018improving, mi2022towards, mi2023graph}: \textit{``unreadable''}, \textit{``neutral''}, and \textit{``readable''}. The mapping we performed was based on the inherent meaning of the Likert scale for this task: The scores $s \in \{1, 2\}$ were mapped to \textit{``unreadable''}, the score $s = 3$ was mapped as \textit{``neutral''}, and the scores $s \in \{4, 5\}$ as \textit{``readable''}. We did this for both \databw and \datas.

\subsection{Experimental Procedure}
\textbf{\RQ{1}: Generalist Models.} To answer \RQ{1}, we leverage two different kinds of machine learning models on the previously-reported datasets: A state-of-the-art model and a Large Language Model (LLM). Since these are generalist models, they can output a single readability label for a single snippet. In our evaluation scenario, we measure the generalist models predictions for each developer ground-truths.
We adopted the model by Scalabrino \etal \cite{scalabrino2018comprehensive} as state-of-the-art model since recent work showed that it appears to be more correlated to professional developers' judgments \cite{sergeyuk2024reassessing}.

We retrained and tested the model for our different scenario as follows.
Given a developer $d$ for which we want to predict the readability scores, we first removed from the datasets the scores they provided (thus simulating the realistic scenario for which we have no data for $d$). Then, we assign the most frequent developer-specific label given by the other developers to each snippet in the set (ground truth). This allows us to define a suitable training dataset for a generalist model and a given developer $d$, letting the model learn from the most shared developer preferences. Finally, we performed a 10-fold cross-validation for each developer.
With this, we have a model which is trained on common readability preferences (\ie generalist model), and test it on each developers' subjective assessments of code readability (as defined in \RQ{1}).

As for the LLM, we use GPT-4o \cite{hurst2024gpt} as a representative Large Language Model. Since this is an already trained model able to follow user instructions, we directly prompt GPT-4o to predict the readability label of a given snippet (\ie we do not re-train it). Specifically, we adopt the following prompt template:
\begin{tcolorbox}[colback=gray!5,colframe=gray!75!black,title=Prompt template for a generalist model]
You are an expert code readability labeler. \\
Your role is to assign a code readability label. The labels you can assign are: \textit{Unreadable}, \textit{Neutral}, and \textit{Readable}.

Now, assign the code readability label for the following snippet:\\
\texttt{target-snippet}.
\end{tcolorbox}
where \texttt{target-snippet} is the snippet to evaluate. We set the temperature to 0 to limit the variability in the model responses.

Finally, we compare the models described above with a hypothetical optimal generalist model, which we assume can consistently return the most commonly assigned label for each code snippet. This serves as an upper bound of a generalist model.

To evaluate the model predictions for each developer score, we compute precision, recall, and F1-score for all readability labels. Precision is calculated as the ratio of correctly predicted instances for each label to the total predictions for that label. Recall is the ratio of correctly predicted instances for each label to the total actual instances of that label. F1-score, the harmonic mean of precision and recall, balances these two metrics.

\textbf{\RQ{2}: Personalized Models.} To answer \RQ{2}, as previously mentioned, we adopt an LLM for defining a personalized model to predict developer subjective assessments of code readability. The rationale is to tailor the model to developer preferences by providing it with previous assessments that can reflect their singular characteristics.
We adopt GPT-4o \cite{hurst2024gpt} as a representative LLM for the personalized model since GPT-based models have been shown very effective under few-shot/in-context learning settings for code-related tasks \cite{sun2024source, ahmed2024automatic,nashid2023retrieval,ahmed2024studying,ahmed2022few,yang2024evaluation}.

In our setting, we decided to set the number of examples $k$ that we should provide to the model to 3 (three-shot learning), conjecturing that providing such a number of code readability evaluation would be acceptable to most developers.

The first shot selection algorithm we test is \textbf{HV} (Highest Variance examples). We select the three snippets that have the highest readability score variance across developers different from $d$. The idea is to ask the developers' score on examples that appear to be controversial in terms of readability for the developers. In our evaluation, we simulate this algorithm by (i) removing the column related to the evaluations of $d$ from the datasets, (ii) computing the variance of the evaluations, (iii) selecting the three snippets with the highest variance, and (iv) assigning to the them the labels provided for those snippets by $d$.

The limitation of HV is that it might not select an example for each label (\eg it could select three snippets that $d$ judged as \textit{readable}), thus not providing the model with enough information to classify snippets from different classes. This problem is due to the lack of knowledge of what will be the evaluation provided by the developer. To understand the ideal performance of HV, we test a non-realistic version of it, (\textbf{HV$_l$} --- Highest Variance examples by Label). HV$_l$ knows the evaluations provided by the developer $d$ to all the snippets and selects, for each label, the snippet with the highest variance among other developers. Note that this is never possible, in practice, but we can do it in our evaluation because we have the vector of evaluations provided by $d$.
In other words, HV$_l$ is a version of HV that always provides an example for each label.

Finally, for each developer, we test a simple third variant that randomly picks the three snippets that should be evaluated (\textbf{R}).

For each snippet selection algorithm ($SSA$) we define a fixed prompt $p_{SSA}^{d_i}$ for each developer $d_i$. The prompt template we adopted is reported at the end of this section. The three shots contain the retrieved code snippets based on the $SSA$ and the developer specific code readability assessment, while \texttt{target-snippet} represents the code snippet to label based on developer preferences.

We compare each personalized model with the best generalist model from \RQ{1}, which works as a baseline, to better measure the actual contributions of such an approach. We test the models only on the set of snippets that are never chosen as examples by the given $SSA$ for any developer.
We evaluate the model predictions as described in \RQ{1}.
\begin{tcolorbox}[colback=gray!5,colframe=gray!75!black,title=Prompt template for a given $SSA$,width=\linewidth]
You are an expert and personal code readability labeler. \\
Your role is to assign a code readability label based on the already known preferences of the developer. The labels you can assign are: \textit{Unreadable}, \textit{Neutral}, and \textit{Readable}.

Below you find examples of the already known developer preferences.

\texttt{$SSA_1$}; \texttt{$SSA_2$}; \texttt{$SSA_3$}

Now, assign the developer code readability label for the following one:\\
\texttt{target-snippet}.
\label{box:prompt_SSA}
\end{tcolorbox}

\textbf{\RQ{3}: Assessments Consistency.}
To answer \RQ{3}, we conduct a qualitative analysis on the developers code readability assessments. We randomly extracted 384 pairs of snippets $\langle s_x, s_y \rangle$ (with $x \neq y$). We chose such a sample size since this ensures a confidence level of 95\% and an expected margin of error of $\pm$ 5\%.
We annotate the snippets of each pair with the respective evaluations provided by a randomly selected developer $d_i$.
We do this for both \databw and \datas (totaling 768 annotated pairs). 

Two authors were tasked to independently evaluate to what extent the evaluation provided for $s_x$ and $s_y$ of each pair (again, by the same developer) was consistent. We say that a pair of evaluations is consistent if there is at least a rational reason (even if not subjectively acceptable to the evaluators) for explaining it, while it is inconsistent otherwise. In other words, we do not assess the consistency based on our concept of code readability. For example, if $s_x$ and $s_y$ have very similar code readability aspects (\eg reasonable comments, proper indentation, \etc) but the developer assigned conflicting assessments (\eg one rated as \textit{readable} and the other as \textit{unreadable}), we mark the evaluation as \textit{inconsistent}. On the other hand, if there is an objectively observable difference, even if contrasting with our notion of readability (\eg the developer reported that the code with more descriptive identifiers is less readable than the other), we mark the evaluation as \textit{consistent}.

Each annotator independently evaluated their own samples reporting ``yes'' if the assessment given by the developer to the two snippets was consistent and ``no'' otherwise. After the independent evaluation, a third author assessed the cases in which the two original evaluators disagreed (\ie one of them reported ``yes'' while the other assessed ``no'') and provided a third judgment as a tie-breaker.
We report the percentage of the inconsistent assessments we found for each dataset.


\section{Empirical Study Results} \label{sec:results}

\subsection{RQ1: Generalist Models}
\label{rq1}
\begin{table*}[t]
\caption{Generalist models on developers for \datas.}
\centering
\resizebox{0.9\linewidth}{!}{%
\begin{tabular}{l|rrr|rrr|rrr}
\rowcolor{black}
& \multicolumn{3}{c}{\textcolor{white}{\textbf{Scalabrino \etal \cite{scalabrino2018comprehensive}}}} 
& \multicolumn{3}{c}{\textcolor{white}{\textbf{GPT-4o}}} 
& \multicolumn{3}{c}{\textcolor{white}{\textbf{Optimal Generalist Model}}}\\
\midrule
\rowcolor{gray!20}
  & \textbf{Precision} & \textbf{Recall} & \textbf{F1-score}
  & \textbf{Precision} & \textbf{Recall} & \textbf{F1-score}
  & \textbf{Precision} & \textbf{Recall} & \textbf{F1-score} \\ \midrule
Unreadable & 0.38 & 0.17 & 0.21 & 0.33 & 0.13 & 0.14 & 0.47 & 0.42 & 0.38\\
Neutral    & 0.41 & 0.16 & 0.22 & 0.31 & 0.66 & 0.40 & 0.53 & 0.44 & 0.46 \\
Readable   & 0.62 & 0.92 & 0.73 & 0.67 & 0.43 & 0.51 & 0.71 & 0.85 & 0.76 \\
\midrule
Average    & 0.47 & 0.42 & 0.39 & 0.44 & 0.41 & 0.35 & 0.57 & 0.57 & 0.53 \\
\bottomrule
\end{tabular}
}
\label{tab:generalized_on_devs_mean_sc}
\end{table*}

\begin{table*}[t]
\caption{Generalist models on developers on \databw.}
\centering
\resizebox{0.9\linewidth}{!}{%
\begin{tabular}{l|rrr|rrr|rrr}
\rowcolor{black}
& \multicolumn{3}{c}{\textcolor{white}{\textbf{Scalabrino \etal \cite{scalabrino2018comprehensive}}}} 
& \multicolumn{3}{c}{\textcolor{white}{\textbf{GPT-4o}}} 
& \multicolumn{3}{c}{\textcolor{white}{\textbf{Optimal Generalist Model}}}\\
\midrule
\rowcolor{gray!20}
  & \textbf{Precision} & \textbf{Recall} & \textbf{F1-score}
  & \textbf{Precision} & \textbf{Recall} & \textbf{F1-score}
  & \textbf{Precision} & \textbf{Recall} & \textbf{F1-score} \\ \midrule
Unreadable & 0.48 & 0.44 & 0.40 & 0.43 & 0.16 & 0.20 & 0.52 & 0.49 & 0.45 \\
Neutral    & 0.00 & 0.00 & 0.00 & 0.27 & 0.74 & 0.39 & 0.36 & 0.12 & 0.18 \\
Readable   & 0.40 & 0.80 & 0.49 & 0.39 & 0.15 & 0.20 & 0.46 & 0.78 & 0.53 \\
\midrule
Average    & 0.29 & 0.41 & 0.30 & 0.36 & 0.35 & 0.26 & 0.44 & 0.46 & 0.39\\
\bottomrule
\end{tabular}
 }
\label{tab:generalized_on_devs_mean_bw}
\end{table*}

\tabref{tab:generalized_on_devs_mean_sc} reports the performance of the generalist models on \datas, while \tabref{tab:generalized_on_devs_mean_bw} reports the performance on the \databw. Due to space limitations, both tables present results in terms of the mean across all developers. Individual developer performances are available in the replication package \cite{replicationpackage}.

Starting with \datas (\tabref{tab:generalized_on_devs_mean_sc}), we observe that all models, including the hypothetical Optimal Generalist Model, achieve relatively poor performance in predicting developers subjective assessments of code readability. This shows a limitation for an ad-hoc practical usage, which motivates the need for personalized approaches as explored in \RQ{2}.
For the ``\textit{unreadable}'' label, which is valuable in practical scenarios, the Optimal Generalist Model achieves a recall of 0.42 and an F1-score of 0.38, indicating that it struggles to identify unreadable code. This shows the divergent perceptions among developers on what constitutes ``\textit{unreadable}'' code, even with an ideal model that leverages the most frequently assigned label for each snippet.

Comparing the model by Scalabrino \etal \cite{scalabrino2018comprehensive} and GPT-4o, we find that the former outperforms the latter. Notably, for the ``\textit{unreadable}'' label, the feature-based model achieves an F1-score of 0.21, compared to the much lower score obtained by GPT-4o (0.14). This trend of higher performance extends across other labels as well, with the feature-based model reaching an overall accuracy of 58\% and an average F1-score of 0.39, whereas GPT-4o achieves only 44\% accuracy and a macro average F1-score of 0.35.

Turning to the \databw (\tabref{tab:generalized_on_devs_mean_bw}), we observe a similar trend as with \datas, where all models demonstrate limited effectiveness in predicting developers' subjective readability assessments.
The Optimal Generalist Model still shows limited effectiveness in predicting the ``\textit{unreadable}'' label, achieving a recall of 0.49 and an F1-score of 0.45, suggesting that developer assessments for ``\textit{unreadable}'' code are not consistently aligned.

Comparing the model by Scalabrino \etal \cite{scalabrino2018comprehensive} to GPT-4o \databw, we find again that the feature-based model generally performs better, particularly for the ``\textit{unreadable}'' label, where it achieves an F1-score of 0.40 compared to GPT-4o 0.20. The model by Scalabrino \etal \cite{scalabrino2018comprehensive} scores 0.00 across all metrics for the ``\textit{neutral}'' label, likely due to the very low number of instances labeled as neutral in the dataset (only nine). This imbalance makes it challenging for the model to learn and predict this label effectively, further impacting its overall performance.

\begin{tcolorbox}[colback=gray!20,colframe=black,title=Answer to \RQ{1}]
Generalist models are not effective for personalized code readability assessments. Surprisingly, GPT-4o achieves lower performance than a significantly smaller, feature-based model tailored for readability predictions.
\end{tcolorbox}

\begin{table}
\caption{Feat model vs GPT4o 3-shots on \datas.}
\centering
\resizebox{\linewidth}{!}{%
\begin{tabular}{l|rrr|rrr}
\rowcolor{black}
& \multicolumn{3}{c}{\textcolor{white}{\textbf{Scalabrino \etal \cite{scalabrino2018comprehensive}}}} 
& \multicolumn{3}{c}{\textcolor{white}{\textbf{GPT-4o - HV}}} \\
\midrule
\rowcolor{gray!20}
& \textbf{Precision} & \textbf{Recall} & \textbf{F1-score} & \textbf{Precision} & \textbf{Recall} & \textbf{F1-score}\\
\midrule
Unreadable & 0.38 & 0.17 & 0.21 & 0.50 & 0.21 & 0.21 \\
Neutral    & 0.41 & 0.16 & 0.22 & 0.29 & 0.36 & 0.31 \\
Readable   & 0.62 & 0.92 & 0.73 & 0.62 & 0.64 & 0.62 \\
\midrule
Average    & 0.47 & 0.42 & 0.39 & 0.47 & 0.41 & 0.38 \\
\bottomrule

\toprule
\rowcolor{black}
& \multicolumn{3}{c}{\textcolor{white}{\textbf{Scalabrino \etal \cite{scalabrino2018comprehensive}}}} 
& \multicolumn{3}{c}{\textcolor{white}{\textbf{GPT-4o - HV$_l$}}} \\
\midrule
\rowcolor{gray!20}
& \textbf{Precision} & \textbf{Recall} & \textbf{F1-score} & \textbf{Precision} & \textbf{Recall} & \textbf{F1-score}\\
Unreadable & 0.38 & 0.17 & 0.21 & 0.34 & 0.15 & 0.19 \\
Neutral    & 0.42 & 0.16 & 0.22 & 0.32 & 0.44 & 0.34 \\
Readable   & 0.62 & 0.92 & 0.73 & 0.65 & 0.63 & 0.62 \\
\midrule
Average    & 0.47 & 0.42 & 0.39 & 0.44 & 0.41 & 0.38 \\
\bottomrule

\toprule
\rowcolor{black}
& \multicolumn{3}{c}{\textcolor{white}{\textbf{Scalabrino \etal \cite{scalabrino2018comprehensive}}}} 
& \multicolumn{3}{c}{\textcolor{white}{\textbf{GPT-4o - R}}} \\
\midrule
\rowcolor{gray!20}
& \textbf{Precision} & \textbf{Recall} & \textbf{F1-score} & \textbf{Precision} & \textbf{Recall} & \textbf{F1-score}\\
Unreadable & 0.38 & 0.17 & 0.21 & 0.46 & 0.15 & 0.19 \\
Neutral    & 0.41 & 0.16 & 0.21 & 0.30 & 0.29 & 0.27 \\
Readable   & 0.62 & 0.92 & 0.73 & 0.62 & 0.75 & 0.66 \\
\midrule    
Average    & 0.47 & 0.42 & 0.39 & 0.46 & 0.39 & 0.38 \\
\bottomrule
\end{tabular}
}
\label{tab:settings_sc}
\end{table}

\begin{table}
\centering
\caption{Feat model vs GPT4o 3-shots on \databw.}
\resizebox{\linewidth}{!}{%
\begin{tabular}{l|rrr|rrr}
\rowcolor{black}
& \multicolumn{3}{c}{\textcolor{white}{\textbf{Scalabrino \etal \cite{scalabrino2018comprehensive}}}} 
& \multicolumn{3}{c}{\textcolor{white}{\textbf{GPT-4o - HV}}} \\
\midrule
\rowcolor{gray!20}
& \textbf{Precision} & \textbf{Recall} & \textbf{F1-score} & \textbf{Precision} & \textbf{Recall} & \textbf{F1-score}\\
\midrule
Unreadable & 0.47 & 0.54 & 0.47 & 0.57 & 0.19 & 0.23 \\
Neutral    & 0.00 & 0.00 & 0.00 & 0.30 & 0.31 & 0.27 \\
Readable   & 0.55 & 0.84 & 0.64 & 0.51 & 0.63 & 0.53 \\
\midrule
Average    & 0.34 & 0.46 & 0.37 & 0.46 & 0.38 & 0.34 \\
\bottomrule

\toprule
\rowcolor{black}
& \multicolumn{3}{c}{\textcolor{white}{\textbf{Scalabrino \etal \cite{scalabrino2018comprehensive}}}} 
& \multicolumn{3}{c}{\textcolor{white}{\textbf{GPT-4o - HV$_l$}}} \\
\midrule
\rowcolor{gray!20}
& \textbf{Precision} & \textbf{Recall} & \textbf{F1-score} & \textbf{Precision} & \textbf{Recall} & \textbf{F1-score}\\
Unreadable & 0.46 & 0.53 & 0.46 & 0.59 & 0.13 & 0.20 \\
Neutral    & 0.00 & 0.00 & 0.00 & 0.29 & 0.39 & 0.31 \\
Readable   & 0.55 & 0.84 & 0.64 & 0.51 & 0.65 & 0.55 \\
\midrule                                             
Average    & 0.34 & 0.46 & 0.37 & 0.46 & 0.39 & 0.35 \\
\bottomrule

\toprule
\rowcolor{black}
& \multicolumn{3}{c}{\textcolor{white}{\textbf{Scalabrino \etal \cite{scalabrino2018comprehensive}}}} 
& \multicolumn{3}{c}{\textcolor{white}{\textbf{GPT-4o - R}}} \\
\midrule
\rowcolor{gray!20}
& \textbf{Precision} & \textbf{Recall} & \textbf{F1-score} & \textbf{Precision} & \textbf{Recall} & \textbf{F1-score}\\
Unreadable & 0.46 & 0.53 & 0.46 & 0.53 & 0.17 & 0.23 \\
Neutral    & 0.00 & 0.00 & 0.00 & 0.29 & 0.42 & 0.31 \\
Readable   & 0.55 & 0.84 & 0.64 & 0.51 & 0.56 & 0.50 \\
\midrule    
Average  & 0.34 & 0.46 & 0.37 & 0.45 & 0.38 & 0.35 \\
\bottomrule
\end{tabular}
}
\label{tab:settings_bw}
\end{table}

\subsection{\RQ{2}: Personalized Models}
We report the results of the personalized models compared with those of the best-performing generalist model (\ie the one by Scalabrino \etal \cite{scalabrino2018comprehensive}) in \tabref{tab:settings_sc} and \tabref{tab:settings_bw}, for \datas and \databw respectively.

Starting with \datas (\tabref{tab:settings_sc}), we observe that the personalized model variants using GPT-4o with different 3-shots snippet selection algorithms (HV, HV$_l$, and R) offer mixed results in comparison to the generalist model. Surprisingly, none of the personalized variants (HV, HV$_l$, or R) achieves consistently higher performance across all metrics and labels, albeit some improvements are evident in specific areas.

Regarding the \textit{``unreadable''} label, GPT-4o with the HV setting (high-variance examples) shows a moderate increase in precision (0.50) compared to the model by Scalabrino \etal \cite{scalabrino2018comprehensive} (0.38). However, it demonstrates a lower recall, 0.21 for HV against the 0.17 of the baseline, leading to a comparable F1-score. This means that GPT-4o HV provides less false positives (unreadable snippets), but misses many actual \textit{``unreadable''} snippets.

In contrast, GPT-4o with the HV$_l$ SSA, which selects one example for each readability label, achieves lower F1-score (0.19) than the baseline (0.21). The same happens for GPT-4o with the R selection algorithm, which randomly select the shots.
Moving to the \textit{``neutral''} and \textit{``readable''} labels, the baseline maintains a clear advantage in F1-score. In details, the model by Scalabrino \etal \cite{scalabrino2018comprehensive} achieves an F1-score of 0.73 on the \textit{``readable''} label, while the highest performance among the personalized models (GPT-4o R) reaches only 0.66, showing that the generalist feature-based model achieves better performance in predicting developers preferences.

As for the dataset \databw (\tabref{tab:settings_bw}), we see a similar pattern to that observed in \datas, with none of the personalized settings consistently outperforming the feature-based model across all metrics and labels. About the \textit{``unreadable''} label, GPT-4o with the HV selection algorithm achieves a higher precision (0.57) than the baseline (0.47), demonstrating a stronger ability to correctly identify unreadable snippets when predicted as such. However, GPT-4o HV shows a low recall of 0.19, compared to the baseline (0.54), leading to an F1-score of only 0.23, which is much lower than the one achieved by the generalist model (0.47). This confirms that while GPT-4o HV is more precise, it fails to predict a significant amount of \textit{``unreadable''} snippets instances.
Regarding the \textit{readable''} labels, the baseline model still holds performance advantage. For the \textit{``readable''} label, Scalabrino's model achieves an F1-score of 0.64, while the highest F1-score among the personalized models reaches only 0.55 (GPT-4o HV$_l$).

\begin{figure*}[t]
    \centering
    \includegraphics[width=0.8\linewidth]{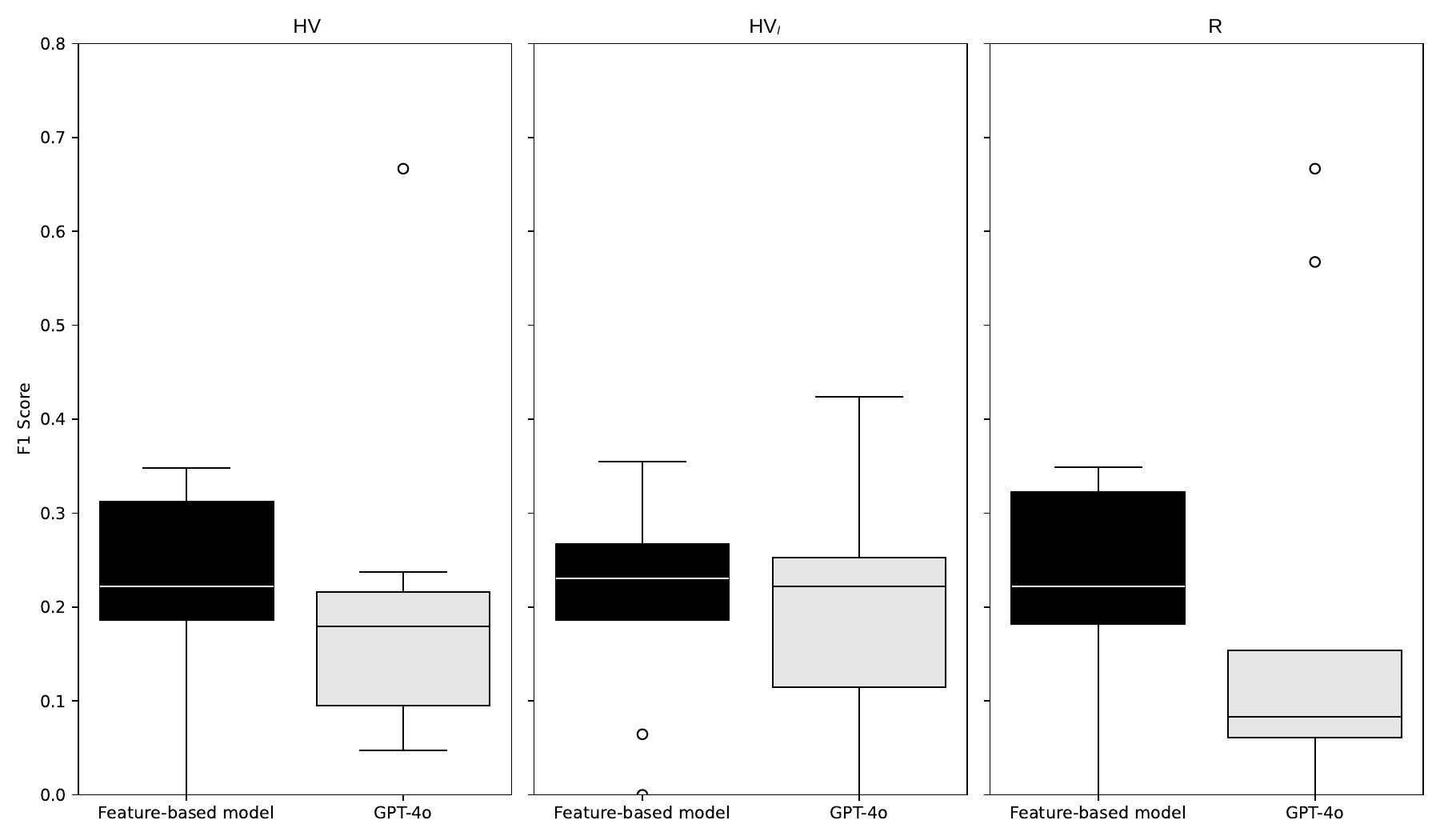}
    \caption{F1-score boxplot for \textit{unreadable} on \datas.}
    \label{fig:unreadable_scalab_plots}
\end{figure*}

\begin{figure*}[h]
    \centering
    \includegraphics[width=0.8\linewidth]{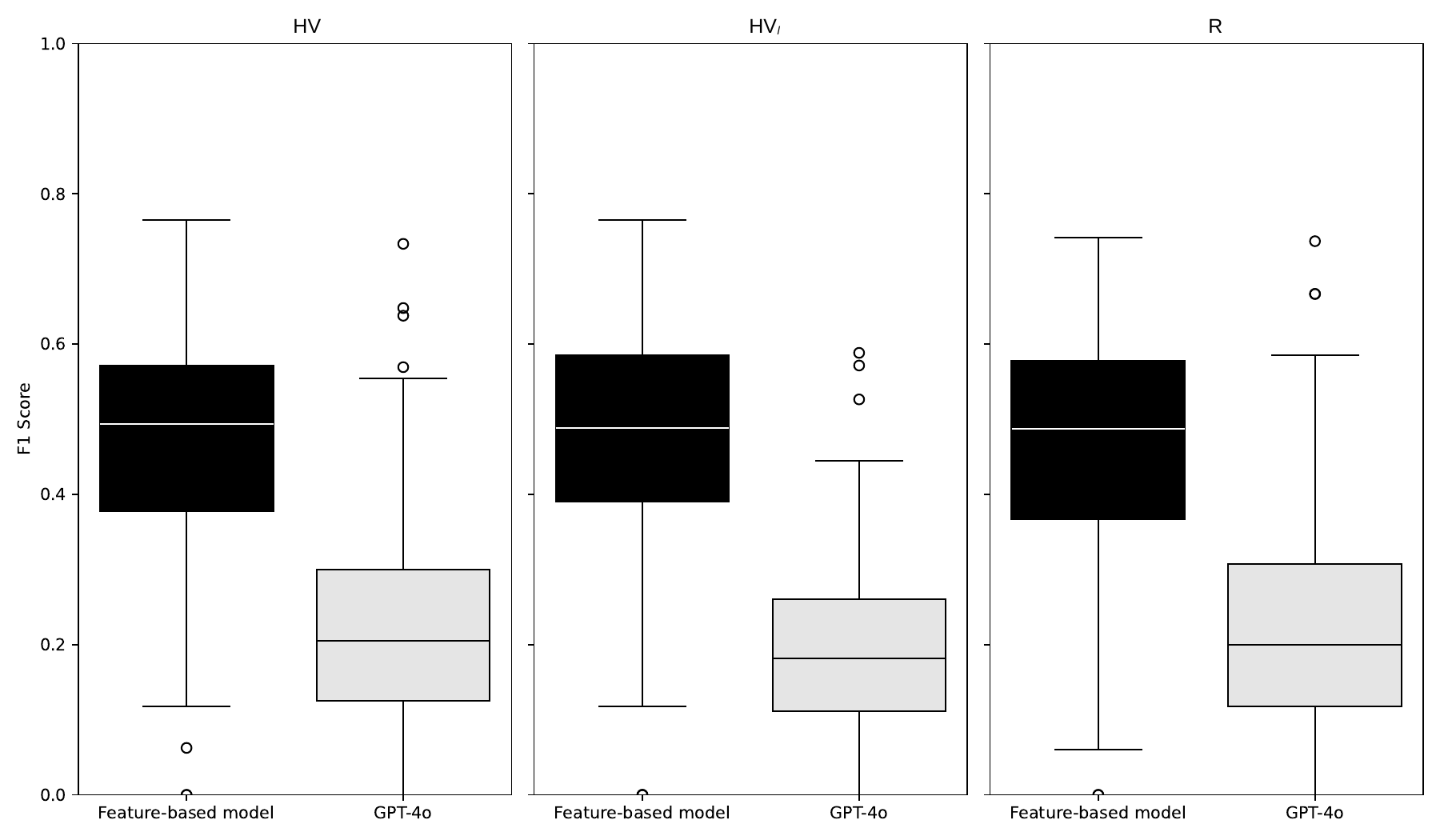}
    \caption{F1-score boxplot for \textit{unreadable} \databw.}
    \label{fig:unreadable_bw_plots}
\end{figure*}
\textbf{Specific on Unreadable.}
In \figref{fig:unreadable_scalab_plots} and \figref{fig:unreadable_bw_plots}, we report the distribution of F1-scores for each developer on the \textit{``unreadable''} label for both \datas and \databw, which represents the most important label in practical scenarios. We compare the feature-based model with the three variants of personalized code readability assessment model.

On \datas, the model by Scalabrino \etal \cite{scalabrino2018comprehensive} consistently shows higher median F1-scores and tighter distributions compared to all GPT-4o settings. In HV, GPT-4o shows a lower median F1-score and greater variability, showing less consistent F1-score than the baseline. In HV$_l$, GPT-4o again shows lower median and a wider range of F1-scores. Lastly, in R (random examples), GPT-4o shows the lowest median F1-score, with scores tightly localized at lower F1-score.

The trends are similar on \databw, but the gap between the baseline and GPT-4o is even more pronounced, providing stronger evidence of the superior feature-based model performance. The latter consistently achieves higher median F1-scores with tighter distributions across all settings. It is worth noting that GPT-4o distributions contain several outliers in HV and HV$_l$, where some developers F1-scores vary widely, which suggest that the model works for some developers. 

\begin{tcolorbox}[colback=gray!20,colframe=black,title=Answer to \RQ{2}]
Personalized LLM-based code readability models are less effective than a state-of-the-art generalist model.
\end{tcolorbox}

\subsection{RQ3: Consistency Analysis}
The manual analysis of the developers code readability assessments revealed a moderate level of inconsistency in evaluations for both \databw and \datas datasets. Among the 384 samples in each dataset, the authors initially disagreed on 23 samples (approximately 6\%) for \databw and 59 samples (approximately 15\%) for \datas. To address these discrepancies, a third author acted as a tiebreaker, carefully reviewing and resolving the disagreements.

As a result, we observed 121 cases (approximately 32\%) of inconsistent evaluations in \databw and 90 cases (approximately 23\%) of inconsistent evaluations in \datas. These results suggest that while developers generally demonstrate consistency in their readability assessments, notable inconsistencies remain, particularly in \databw.

\begin{figure}
    \centering
    \includegraphics[width=\linewidth]{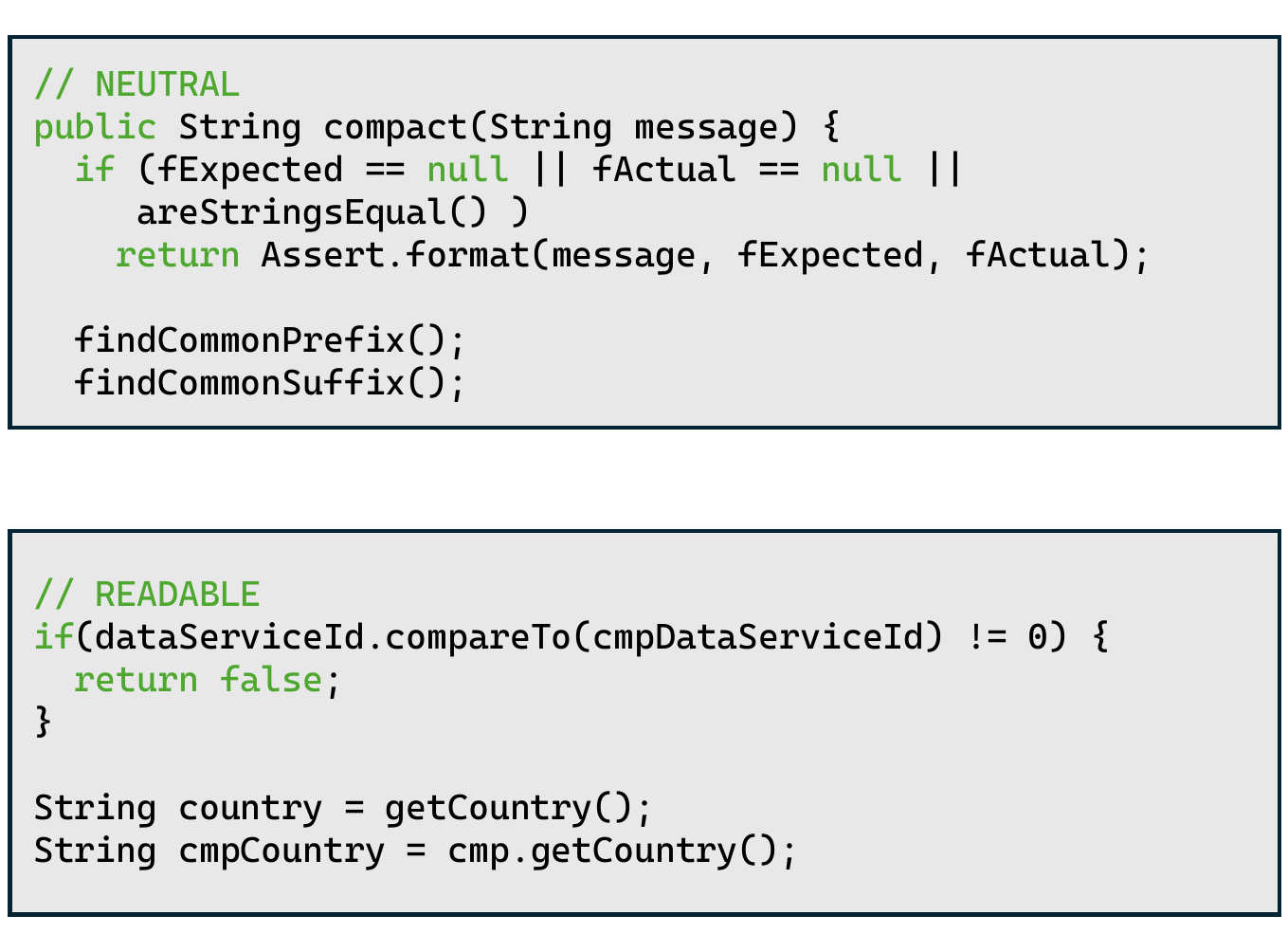}
    \caption{Example of inconsistent evaluation from \databw.}
    \label{fig:results:inconsistent}
\end{figure}

We report in \figref{fig:results:inconsistent} an example of inconsistent evaluation from \databw. The developer annotated the snippet on the left as \textit{neutral} and the one on the right as \textit{readable}. However, the we could not find any possible explanation for the different evaluation since the two are very similar even in terms of their structure.

\begin{tcolorbox}[colback=gray!20,colframe=black,title=Answer to \RQ{3}]
Developers code readability assessments show moderate consistency, with 32\% inconsistency in \databw and 23\% in \datas.
\end{tcolorbox}

\section{Discussion} 
\label{sec:discussion}
We report below the lessons learned that the reader can take away from this study.
One of the main results we obtained is that generalist models achieve bad performances on the task of predicting developer-specific code readability perception. An ideal model only achieves 0.53 F1-score on \datas and 0.39 F1-score on \databw.

\textbf{Lesson Learned 1.} \textit{The definition of personalized code readability prediction models is worth future investigation.}

We observed that state-of-the-art generalist models work very well even in the personalized code readability prediction task. As we observed in \RQ{1}, the accuracy obtained by such models is close to the upper-bound that can be theoretically achieved in such a task by an optimal model.

\textbf{Lesson Learned 2.} \textit{There is limited room for improvement of generalist code readability assessment models.}

Surprisingly, we observed that a feature-based model achieves better results than an LLM-based model (both generalist and personalized). In particular, LLMs seem not to have a good embedded understanding of what \textit{readable} or \textit{unreadable} code mean. It is worth noting that in our experiment we did not use the learning base constituted by the evaluation of other developers for LLMs (while we did for the feature-based model). This could explain the difference we observed. Future work could try to fine-tune such models with such \textit{generic} evaluations, to provide the model with a baseline concept of code readability before adding specific examples tailored on the readability perception of the developer at hand.

\textbf{Lesson Learned 3.} \textit{Fine-tuning could be a viable option to improve the performance of LLMs for code readability prediction, specifically for the personalized models.}

Finally, our manual investigation of the two datasets we adopted in our experiment clearly shows that the annotators provided self-contradictory evaluations. This is mostly due to the simple procedure adopted in previous work for defining the datasets, which consisted in asking developers to self-declare their perceived readability. While such a procedure is good enough for defining generalist models (the noise due to self-contradictions does not significantly affect the average score assigned to the snippet), it becomes not reliable for defining a personalized model. 

\textbf{Lesson Learned 4.} \textit{Future work should aim at using more principled procedures for defining code readability datasets, \eg based on the time needed to read the code \cite{hofmeister2019shorter}.}

\section{Threats to Validity} \label{sec:threats}
\textbf{Threats to construct validity} mainly pertain to the mapping of Likert scale scores into labels (readable, neutral, unreadable) that may not accurately reflect meaningful readability distinctions for all developers. A snippet rated ``2'' by one developer might be considered ``neutral'' by another. However, we believe that our methodology for mapping the scores into labels in inherently more reliable than the one used for generalist models because it relies on the intuitive interpretation of the Likert scale itself (1 and 2 are related to a negative score, 4 and 5 are related to a positive score, while 3 is neutral).
In addition, developers may use the same score for different reasons (\eg one developer might rate a snippet as ``5'' because it is concise, while another rates it ``5'' because it has detailed comments). 

\textbf{Threats to internal validity} concerned the experimental choices that might have affected the results. A first concern is related to the dataset used to answer our RQs (that, as we show in \RQ{3}, is not entirely reliable). Besides, we only considered a scenario in which only three developer-specific examples are provided to the model. It is possible that the negative results obtained in this paper is due to the fact that such a small number of examples might not in principle allow the model to fully capture the nuances of individual preferences.
Also, it is possible that the results of \RQ{3} are somewhat due to subjective and arbitrary assessment by the annotators. We mitigated this threat by using a rigorous qualitative analysis process involving three annotators, as described in \secref{sec:design}.

\textbf{Threats to external validity} pertain the generalizability of the results to other contexts. First, we only focused on Java code. The results might not generalize to other programming languages. 
Also, the settings evaluated in the study may not entirely represent the complexity of real software development environments. It is possible that the personalized model we defined achieves completely different results on developers with a different background. It is worth noting, however, that our experiment indirectly involved 129 developers, which is a quite large sample.
Finally, we only tried our personalized procedure with a specific LLM, \ie GPT-4o. The results might not generalize to other LLMs.

\section{Related Work}
\label{sec:related}
Sergeyuk \etal \cite{sergeyuk2024reassessing} assessed the alignment of existing (generalist) code readability assessment models with developers' views of LLM-generated code. Their study involved (i) creating a dataset of LLM-generated Java code snippets, (ii) applying code readability assessment models to evaluate snippets, (iii) identifying the readability dimensions preferred by developers using the retrieval grid technique, and (iv) comparing model evaluations with human evaluations.
The research identified 12 key dimensions that influence code readability and revealed a weak correlation between existing code readability assessment models and developer ratings. These findings highlight the need to develop more accurate and developer-aligned code readability assessment models to better support software development. 

Sergeyuk \etal \cite{sergeyuk2024assessing} further investigated the consistency of developer assessments and the key aspects driving their evaluations. By surveying 10 Java developers with similar professional backgrounds, they evaluated 30 LLM-generated Java code snippets across 12 readability dimensions. Their findings showed moderate to good agreement among developers (Intraclass Correlation Coefficient = 0.78) and highlighted dimensions such as code length, goal clarity, and naming clarity as strongly correlated with readability. These results emphasize the potential to align LLM-generated outputs with developers' notions of readability by focusing on stable, consensus-based metrics.

While these studies provide significant insights, there are notable gaps yet to be addressed. Sergeyuk \etal \cite{sergeyuk2024reassessing} primarily focus on broad dimensions of readability and shared developer perspectives, without delving into the subjective nature of code readability, specifically how individual developers perceive and evaluate code. Developers bring their experiences, preferences, and cognitive styles to the task of reading and understanding code, which combined datasets and generalized assessments fail to capture. 
Our study addressed these limitations by proposing a personalized code readability assessment approach. 

Previous work already tried to tackle the problem of personalizing recommender systems for other software engineering tasks.
Allamanis \cite{allamanis2014learning} introduced the framework NATURALIZE for learning the stylistic conventions of a codebase and suggesting revisions to ensure consistency. By applying statistical natural language processing to source code, NATURALIZE achieved high accuracy in identifier naming and formatting, with practical applications in development and code review. 

Research conducted in an industry context showed that developer- and application-specific models outperform general models trained on entire codebases. However, temporal changes in codebases have minimal effects on the performance of language models \cite{saraiva2015products}. These outcomes suggest that adapting models to individual developers or projects leads to better results, reinforcing the idea that customization increases the effectiveness of models in software development activities.

\section{Conclusion} \label{sec:conclusion}
We presented a study in which we aimed at understanding whether we can define a personalized code readability prediction model, \ie a model that capture the specific developers' notion of code readability. The results of our investigation clearly show that we are not there yet. Specifically, we observed that (i) generalist models are not sufficient to perform such a task, (ii) even modern and promising technologies, like LLMs, do not achieve satisfactory results, and (iii) currently-available code readability datasets have substantial limitations.
Future work should take up the challenge and provide more reliable datasets and more accurate personalized code readability prediction models.

\section{Data Availability}
We publicly release our replication package \cite{replicationpackage}, in which we provide our datasets, the scripts for building and everything needed to replicate all the results of our experiment.

\section*{Acknowledgments}
This publication is part of the project PNRR-NGEU which has received funding from the MUR – DM 118/2023.
This work has been partially supported by the European Union - NextGenerationEU through the Italian Ministry of University and Research, Projects PRIN 2022 ``DevProDev: Profiling Software Developers for Developer-Centered Recommender Systems'', grant n. 2022S49T4W, CUP: H53D23003610001.

\bibliographystyle{IEEEtranS}
\bibliography{main}

\end{document}